%% file: Correction.tex
\documentclass[a4paper,colorlinks=true,allcolors=ThemeColor,hyperfootnotes=false,11pt]{article}
\usepackage[margin=1in]{geometry}
\usepackage[utf8]{inputenc}
\usepackage[T1]{fontenc}
\usepackage{lmodern,textcomp} 
\usepackage[american]{babel}
\usepackage{microtype}
\usepackage{bookmark}
\usepackage[onehalfspacing]{setspace}
\usepackage[bottom]{footmisc}
\usepackage{graphicx}
\usepackage{rotating}
\usepackage[position=top]{subfig}
\usepackage[font=large]{caption}
\usepackage{pdflscape}
\usepackage{longtable}
\usepackage{threeparttable}
\usepackage{adjustbox}
\usepackage{multirow}
\usepackage{morefloats}
\usepackage{float}
\usepackage{boldline}
\usepackage{tabularx,booktabs}
\usepackage{rotating}
\usepackage{makecell}
\usepackage{caption}

\usepackage{pifont}
\newcommand{\cmark}{\ding{51}}%
\usepackage{amsmath,amsfonts,amsthm,amssymb}
\usepackage{bbm}
\usepackage[titletoc,title]{appendix}
\usepackage{csquotes}
\usepackage[backend=biber,authordate,uniquename=false,uniquelist=false,url=false,doi=false,isbn=false]{biblatex-chicago}
\AtEveryBibitem{
    \clearfield{urlyear}
    \clearfield{urlmonth}
    \clearfield{month}
    \clearfield{day}
    \clearfield{note}
    \clearlist{language}
}
\DeclareNosort{
  \nosort{type_names}{\regexp{[\x{2bf}\x{2018}]}}
}
\DeclareNoinit{
  \noinit{\regexp{[\x{2bf}\x{2018}]}}
}
\usepackage{xcolor}
\definecolor{ThemeColor}{RGB}{34,110,147} 
\usepackage{zref-xr,zref-user}
\usepackage{nameref}
\zxrsetup{toltxlabel}
\zexternaldocument*{CorrectionApp}
\usepackage{pdfpages}
\usepackage[normalem]{ulem}
\usepackage{url}
\usepackage{enumitem}
\setlist{nosep}

\newtheorem*{pred*}{Prediction}

\makeatletter
\def\th@plain{%
  \thm@notefont{}
  \itshape 
}
\def\th@definition{%
  \thm@notefont{}
  \normalfont 
}
\makeatother

\urldef{\PAP}\url{https://osf.io/tgyc5}
\newcommand{\starnote}{Significance levels: * 10\%, ** 5\%, and *** 1\%}
\newcommand{\sdnote}{Standard errors in parentheses are clustered at the individual level}

\usepackage{fancyhdr}
\fancypagestyle{PAP}{%
\fancyhf{}
\fancyhead[L]{Pre-analysis plan}
\fancyfoot[L]{\thepage}
}%
\fancypagestyle{EnglishTrans}{%
\fancyhf{}
\fancyhead[L]{Experimental instructions: English translation}
\fancyfoot[L]{\thepage}
}%

\title{Gender Differences in the Cost of Corrections in Group Work}

\author{Yuki Takahashi\thanks{Department of Economics, University of Bologna. Email: \href{mailto:yuki.takahashi2@unibo.it}{\texttt{yuki.takahashi2@unibo.it}}.
I am grateful to Maria Bigoni, Siri Isaksson, Bertil Tungodden, Laura Anderlucci, and Natalia Montinari whose feedback was essential for this project. I am also grateful to participants of the experiment for their participation and cooperation. Sonia Bhalotra, Francesca Cassanelli, Alessandro Castagnetti, Mónica Costa-Dias, Valeria Ferraro, Lenka Fiala, Ria Granzier-Nakajima, Silvia Griselda, Annalisa Loviglio, Yoko Okuyama, Monika Pompeo, Øivind Schøyen, Vincenzo Scrutinio, Erik Ø. Sørensen, Ludovica Spinola, Florian Zimmermann, and PhD students at the NHH and the University of Bologna all provided many helpful comments. This paper also benefited from participants' comments at the Annual Southern PhD Economics Conference, ESA Conference, FROGEE Workshop, Gender Gaps Conference, PhD-EVS, Stanford Institute for Theoretical Economics, TIBER Symposium, Warwick Economics PhD Conference, WEAI Conference, Webinar in Gender and Family Economics, and seminars at Ca' Foscari University, Catholic University of Brasília, the NHH, Tilburg University, the University of Bologna, and the University of Copenhagen. Ceren Ay, Tommaso Batistoni, Philipp Chapkovski, Sebastian Fest, Christian König genannt Kersting, and oTree help \& discussion group kindly answered my questions about oTree programming; in particular, my puzzle code was heavily based on Christian's code. Michela Boldrini and Boon Han Koh conducted the quasi-laboratory experiments ahead of me and kindly answered my questions about their implementations. Lorenzo Golinelli provided excellent technical and administrative assistance. This study was pre-registered with the OSF registry (\PAP) and approved by the IRB at the University of Bologna (\#262643).
}
\\\\\href{https://yukitakahashi1.github.io/files/Correction.pdf}{Click here for the latest version}
}

\begin{document}
\begin{singlespace}
\maketitle
\begin{abstract}
\noindent
Corrections among colleagues are an integral part of group work, but people may take corrections as personal criticism, especially corrections by women. I study whether people dislike collaborating with someone who corrects them and more so when that person is a woman. People, including those with high productivity, are less willing to collaborate with a person who has corrected them even if the correction improves group performance. Yet, people respond to corrections by women as negatively as by men. These findings suggest that although women do not face a higher hurdle, correcting colleagues is costly and reduces group efficiency.
\end{abstract}
\end{singlespace}

\newpage
\clearpage
\section{Introduction}
Receiving corrections from colleagues is an integral part of group work. Consider academic research. From the development of ideas to the writing up of the final draft, researchers discuss their research project with their colleagues, receive criticisms, and refine the ideas and the analysis. However, people may take the corrections personally. Imagine a researcher presents their paper for which they spent several years, and someone points out a possible flaw in their identification assumption or their experimental design. Since the validity of identification assumptions and experimental designs are debatable, they may take it as a personal criticism.

Those people may express their discomfort in some way, and a least aggressive way to do so is not to collaborate. However, not being invited for collaboration could be detrimental to one's career success because having collaborations is essential in academia, where people co-author the majority of papers \parencite{jones_rise_2021,wuchty_increasing_2007}.

Women's corrections may receive stronger negative reactions because people often use double standards for women and men. Evidence suggests that men undervalue women when they criticize them \parencite{sinclair_motivated_2000} and that people punish women more harshly when they make mistakes \parencite{sarsons_interpreting_2019} and commit misconduct \parencite{egan_when_2021}. If so, women face a higher hurdle in their career success. It is also detrimental to group efficiency as group members cannot fully benefit from female colleagues.

This paper studies whether people dislike collaborating with someone who corrects them and more so when that person is a woman. Answering this question using secondary data poses two challenges. First, group formation is not random and group corrections are endogenous. Second, different corrections are not necessarily comparable to each other.

Thus, I design a quasi-laboratory experiment, a hybrid of physical laboratory and online experiments, where group formation is randomized and define corrections such that researchers can track its quality mathematically. Specifically, participants are allocated to a group of eight and solve one joint task with each group member one by one. Each time participants finish the task, they state whether they would like to collaborate with the group member with whom they have just solved the task for the same task in the next stage, which is the main source of earnings. This gives a strong incentive for participants to select as good a collaborator as possible. The order of the group members with whom participants solve the task is randomized. As a joint task, I use \textcite{isaksson_it_2018}'s number-sliding puzzle, which allows me to calculate an objective measure of each participant's contribution to the joint task as well as to classify each move as good or bad. I define a correction as reversing a group member's move, which is comparable across different participants and can be classified as either good or bad.

I find that participants correctly understand the notion of good and bad moves; that is, the higher your contribution is to solving the puzzle, the more likely it is that you will be asked to join a team. This is in line with what one would expect and validates my experimental design.

Nonetheless, after controlling for the contribution, people are significantly less willing to collaborate with a person who has corrected their moves, even if the corrections move the puzzle closer to the solution. Although it may not be so costly to correct colleagues if only low productivity people respond negatively to corrections, high productivity people also respond negatively to corrections.

Yet, people respond to corrections by women as negatively as by men; although I find suggestive evidence that men respond more negatively to women's good corrections, the finding is not robust. These findings are unlikely to be due to people's belief about differences in women's and men's abilities: women and men contribute equally to the puzzle, and neither women nor men underestimate women's contribution.

This paper primarily relates to studies on gender differences in the contribution of ideas in group work. \textcite{coffman_evidence_2014} finds that women are less likely to contribute their ideas to the group in a male task due to self-stereotyping and \textcite{gallus_shine_2019} find that debiasing their self-stereotyping by giving an award for their high ability increases women’s contribution of their ideas: they put women's idea further ahead without involving open correction of their group member. However, on some occasions, the contribution of ideas has to be made openly, for example, in academic seminars and business meetings. In such cases, group members' response plays an important role in the effectiveness of the intervention. \textcite{coffman_gender_2021} find that group members are less likely to choose women's answers as a group answer in male-typed questions. \textcite{guo_overriding_2020} find that group members correct women’s ideas more often than men’s ideas. \textcite{dupas_gender_2021} find that female economists receive more patronizing and hostile questions during seminars. \textcite{isaksson_it_2018} finds that men are more likely to correct their group member's bad moves in the same puzzle used in my experiment.\footnote{As the puzzle was originally used by \textcite{isaksson_it_2018}.} My paper introduces correction in the contribution of ideas and examines its cost on women and on group efficiency.

More generally, my paper contributes to the literature on gender differences in group work. \textcite{isaksson_it_2018} finds that women under-claim their contribution compared to men in group work despite their equal contribution. \textcite{haynes_it_2013} find similar results. \textcite{sarsons_gender_2021} find that people attribute less credit to a female economist when she co-authors a paper with a male economist(s). \textcite{born_gender_2020} and \textcite{stoddard_strength_2020} find that women are less willing to lead a male-majority group. \textcite{shan_does_2020} finds that female students are more likely to drop out from an introductory economics class when they are assigned to a male-majority study group. \textcite{babcock_gender_2017} find that women are more likely to volunteer and be asked to do non-promotable tasks. My paper promotes our understanding of gender differences in group work.

My paper also speaks to the literature on organizational efficiency. \textcite{bandiera_social_2009} find that managers favor workers who have connections with the managers, which distorts the allocation of talent and reduces firm efficiency as theory predicts \parencite{macleod_optimal_2003,prendergast_favoritism_1996}. \textcite{li_costs_2020} finds that this managers' favoritism not only distorts the optimal allocation of talent but also reduces non-favored workers' performance. \textcite{cullen_old_2021} find that managers favor workers with whom they presumably take a smoking break together. \textcite{fang_gender_2017} find that institutional investors give positively biased evaluations for corporate analysts who graduated from the same university, especially when the analysts are men. \textcite{kennedy_when_2008} find that people tend to view others who disagree with them as biased. \textcite{ronayne_ignoring_2019} find that people often stick to their decisions rather than accepting decisions suggested by people with higher abilities. My finding that people are less willing to collaborate with people who have corrected them can be another source of organizational inefficiency.

\section{Experiment}
\paragraph{Introducing a quasi-laboratory format}
I run the experiment in a quasi-laboratory format where we experimenters connect us to the participants via Zoom throughout the experiment (but turn off participants' camera and microphone except at the beginning of the experiment) and conduct it as we usually do in a physical laboratory, but participants participate remotely using their computers. Online Appendix section \ref{sec:ProsConsQuasiLab} discusses the pros and cons of the quasi-laboratory format relative to physical laboratory and standard online experiments.

\begin{figure}[!htb]
\begin{center}
\caption{Puzzle screen}\label{fig:PuzzleScr}
\includegraphics[width=\columnwidth]{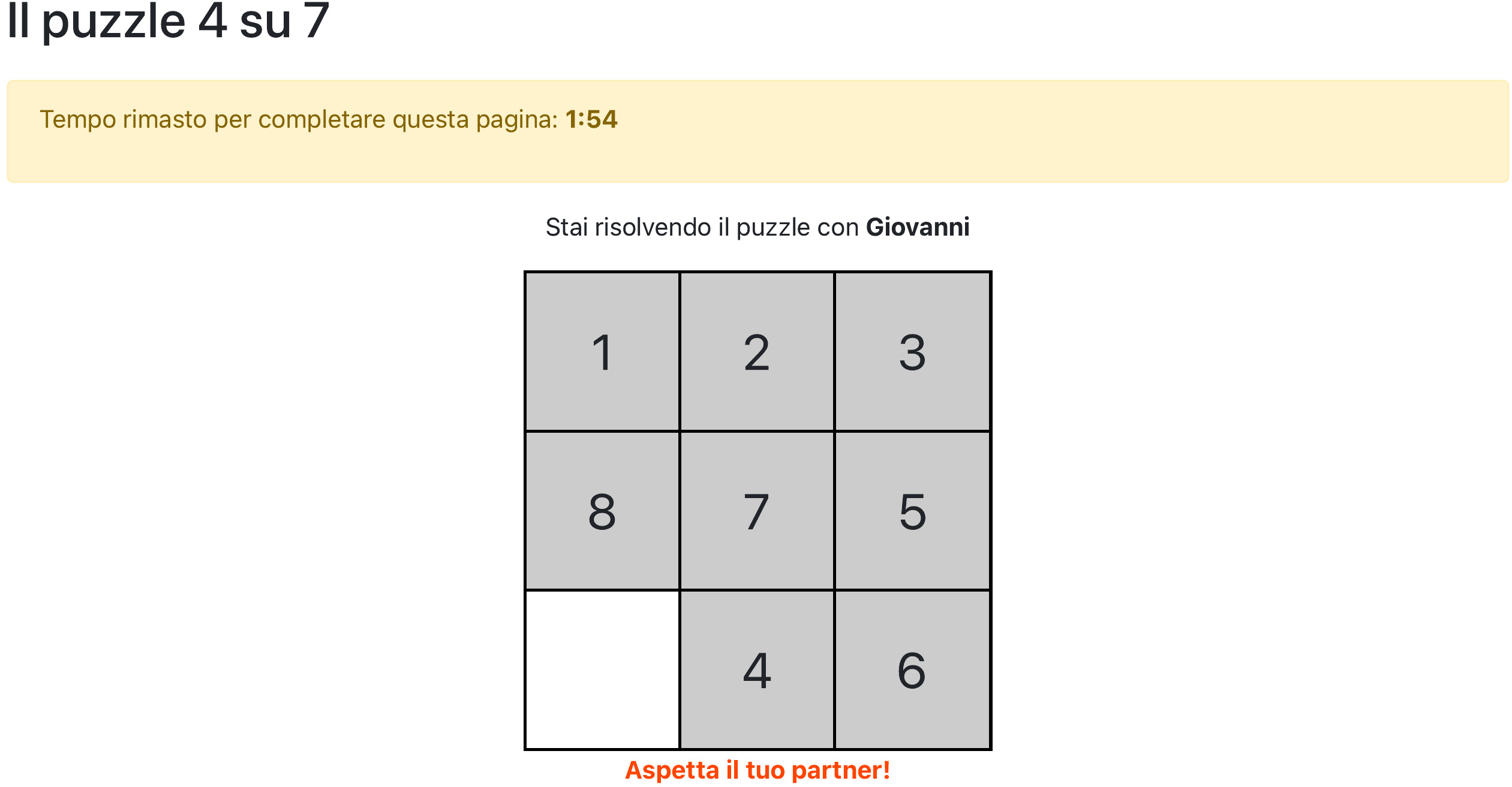}
\end{center}
\vspace{-12pt}
\begin{tablenotes}[flushleft]
\item \footnotesize \textit{Notes:} This shows a sample puzzle screen where a participant is matched with another participant called Giovanni at the 4th round puzzle and waiting for Giovanni to make his move.
\end{tablenotes}
\end{figure}

\paragraph{Group task}
As the group task, I use \textcite{isaksson_it_2018}'s puzzle, a sliding puzzle with eight numbered tiles, which should be placed in numerical order within a 3x3 frame (see Figure \ref{fig:PuzzleScr} for an example). To achieve this goal, participants play in pairs, alternating their moves. This puzzle has nice mathematical properties: I can define the puzzle difficulty and classify a given move as either good or bad by the Breadth-First Search algorithm. From the number of good and bad moves one makes, I can calculate individual contributions to the group task; I measure it by net good moves, the number of good moves minus the number of bad moves an individual makes in a given puzzle.

I can also determine the quality of corrections of different participants objectively and comparably.\footnote{The difficulty is defined as the number of moves away from the solution, a good move is defined as a move that reduces the number of moves away from the solution, and a bad move is defined as a move that increases the number of moves away from the solution.} Further, the puzzle-solving captures an essential characteristic of teamwork in which two or more people work towards the same goal \parencite{isaksson_it_2018}, but the quality of each move and correction is only partially observable to participants (but fully observable to the experimenter).

At each stage of the puzzle, there is only one best strategy which is to make a good move.\footnote{This is conditional on that both players are trying to solve the puzzle; I show in online Appendix \ref{sec:Robustness} that the results are robust to exclusion of puzzles where either player might not be trying to solve the puzzle.} There can be more than one good and bad move, but different good/bad moves are equal. There is no path dependence either: the history of the puzzle moves does not matter.

At the beginning of each part, participants must answer a set of comprehension questions to make sure they understand the instructions.

\subsection{Design and procedure}\label{sec:Design}
\subsubsection*{Registration}
Upon receiving an invitation email to the experiment, participants register for a session they want to participate in and upload their ID documents as well as a signed consent form.\footnote{I recruit a few more participants than I would need for a given session in case some participants would not show up to the session.}

\subsubsection*{Pre-experiment}
On the day and the time of the session they have registered for, participants enter the Zoom waiting room.\footnote{Zoom link is sent with an invitation email; I check that they have indeed registered for a given session before admitting them to the Zoom meeting room.} They receive a link to the virtual room for the experiment and enter their first name, last name, and their email they have used in the registration. They also draw a virtual coin numbered from 1 to 40 without replacement.

Then I admit participants to the Zoom meeting room one by one and rename them by the first name they have just entered. This information is necessary to match up their earnings in this experiment and their payment information stored in the laboratory database, so participants have a strong incentive to provide their true name and email address. If there is more than one participant with the same first name, I add a number after their first name (e.g., Giovanni2).

After admitting all the participants to the Zoom meeting room, I do roll call, a way to reveal participants' gender to other participants without making gender salient \parencite{bordalo_beliefs_2019,coffman_gender_2021}. Specifically, I take attendance by calling each participant's first name one by one and ask her or him to respond via microphone. This process ensures other participants that the called participant's first name corresponds to their gender. If there are more participants than I would need for the session (I need 16 participants), I draw random numbers from 1 to 40 and ask those who drew the coins with the same number to leave.\footnote{I draw with replacement a number from 1 to 40 using Google's random number generator (\url{https://www.google.com/search?q=random+number}). If no participant has a coin with the drawn number, I draw next number until the number of participants is 16. I share my computer screen so that participants see the numbers are actually drawn randomly.} Those who leave the session receive the 2€ show-up fee. Online Appendix Figure \ref{fig:ZoomScr} shows a Zoom screen participants would see during the roll call (the person whose camera is on is the experimenter; participants would see this screen throughout the experiment, but the experimenter's camera may be turned off).

I then read out the instructions about the rules of the experiment and take questions on Zoom. Once participants start the main part, they can communicate with the experimenter only via Zoom's private chat.

\subsubsection*{Part 1: Individual practice stage}
Participants work on the puzzle individually with an incentive (0.2€ for each puzzle they solve). They can solve as many puzzles as possible with increasing difficulty (maximum 15 puzzles) in 4 minutes. This part familiarizes them with the puzzle and provides us with a measure of their ability given by the number of puzzles they solve. After the 4 minutes are over, they receive information on how many puzzles they have solved.

\subsubsection*{Part 2: Collaborator selection stage}
Part 2 contains seven rounds, and participants learn the rules of part 3 before starting part 2. This part is based on \textcite{fisman_gender_2006,fisman_racial_2008}'s speed dating experiments and proceeds as follows: first, participants are allocated to a group of 8 based on their ability similarity as measured in part 1. This is done to reduce ability difference among participants, and participants do not know this grouping criterion.

Second, participants are paired with another randomly chosen participant in the same group and solve one puzzle together by alternating their moves. The participant who makes the first move is drawn at random and both participants know this first-mover selection criterion. If they cannot solve the puzzle within 2 minutes, they finish the puzzle without solving it. Participants are allowed to reverse the paired participant's move.\footnote{Solving the puzzle itself is not incentivized, and thus participants who do not want to collaborate with the paired participant or fear to receive a bad response may not reverse that participant's move even if they think the move is wrong. However, since I am interested in the effect of correction on collaborator selection, participants' \textit{intention} to correct that does not end up as an actual correction does not confound the analysis.} Reversing the partner's move is what I call correction in this paper. Each participant's contribution in a given puzzle is measured by net good moves. Figure \ref{fig:PuzzleScr} shows a sample puzzle screen where a participant is paired with another participant called Giovanni and waiting for Giovanni to make his move. The paired participant's first name is displayed on the computer screen throughout the puzzle and when participants select their collaborator to subtly inform the paired participant's gender.

Once they finish the puzzle, participants state whether they would like to collaborate with the same participant in part 3 (yes/no). At the end of the first round, new pairs are formed, with a perfect stranger matching procedure, so that every participant is paired with each of the other seven members of their group once and only once. In each round, participants solve another puzzle in a pair, then state whether they would like to collaborate with the same participant in part 3. The sequence of puzzles is the same for all pairs in all sessions. The puzzle difficulty is kept the same across the seven rounds. The minimum number of moves to solve the puzzles is set to 8 based on the pilot.

At the end of part 3, participants are paired according to the following algorithm:
\begin{enumerate}
\item For every participant, call it i, I count the number of matches; that is, the number of other participants in the group who were willing to be paired with i and with whom i is willing to collaborate in part 3.
\item I randomly choose one participant.
\item If the chosen participant has only one match, I pair them and let them work together in part 3.
\item If the chosen participant has more than one match, I randomly choose one of the matches.
\item I exclude two participants that have been paired and repeat (1)-(3) until no feasible match is left.
\item If some participants are still left unpaired, I pair them up randomly.
\end{enumerate}

\subsubsection*{Part 3: Group work stage}
The paired participants work together on the puzzles by alternating their moves for 12 minutes and earn 1€ for each puzzle solved. Which participant makes the first move is randomized at each puzzle, and this is told to both participants as in part 2. They can solve as many puzzles as possible with increasing difficulty (maximum 20 puzzles).

\subsubsection*{Post-experiment}
Each participant answers a short questionnaire which consists of (i) the six hostile and benevolent sexism questions used in \textcite{stoddard_strength_2020} with US college students and (ii) their basic demographic information and what they have thought about the experiment.\footnote{I was planning to construct a gender bias measure from the hostile and benevolent sexism questions and use it to show those with higher gender bias respond more negatively to women's corrections. However, people do not respond more negatively to women's corrections and that I could not have enough variation in this gender bias measure, so decided not to report it. See the pre-analysis plan in the online Appendix section \ref{sec:PAP}.} The answer to their demographic information is used to know participants' characteristics as well as casually check whether they have anticipated that the experiment is about gender, for which I do not find any evidence.

After participants answer all the questions, I tell them their earnings and let them leave the virtual room and Zoom. They receive their earnings via PayPal.

\subsection{Implementation}
The experiment was programmed with oTree \parencite{chen_otreeopen-source_2016} and conducted in Italian during November-December 2020. I recruited 464 participants (244 female and 220 male) registered on the Bologna Laboratory for Experiments in Social Science's ORSEE \parencite{greiner_subject_2015} who (i) were students, (ii) were born in Italy, and (iii) had not participated in gender-related experiments before (as far as I could check).\footnote{The laboratory prohibits deception, so no participant has participated in an experiment with deception.} The first two conditions were to reduce noise coming from differences in socio-demographic backgrounds and race or/and ethnicity that may be inferred from participants' first name or/and voice, and the last condition was to reduce experimenter demand effects. The number of participants was determined by a power simulation in the pre-analysis plan to achieve 80\% power.\footnote{This number includes 16 participants from a pilot session run before the pre-registration where the experimental instructions were slightly different. The results are robust to exclusion of these 16 participants.} The experiment is pre-registered with the OSF.\footnote{The pre-registration documents are available at the OSF registry: \PAP. The pre-analysis plan is also in the online Appendix \ref{sec:PAP}.}

I ran 29 sessions with 16 participants each. The average duration of a session was 70 minutes. The average total payment per participant was 11.55€ with the maximum 25€ and the minimum 2€, all including the 2€ show-up fee. Online Appendix Table \ref{tab:Summary} describes participants' characteristics. The table shows that female participants are more likely to major in humanities and male participants are more likely to major in natural sciences and engineering, a tendency observed in most OECD countries \parencite[see, for example,][]{carrell_sex_2010}.\footnote{Individual fixed effects in the analysis control for one's major. However, I do not run heterogeneity analysis by major because major choice is endogenous to one's gender.} Also, most female and male participants are either bachelor or master students (97\% of female and 94\% of male) and the rest are PhD students.

\section{Data description}\label{sec:Data}
I use part 2 data in the analysis as part 2 is where we can observe collaborator selection decisions. I aggregate the move-level data at each puzzle so that we can associate behaviors in the puzzle to the collaborator selection decisions. As shown in online Appendix Figure \ref{fig:PuzzleMoves}, both mixed gender and single-gender groups perform equally well (panel A), about 71\% of the puzzles are solved within a minimum number of moves (panel B, the minimum number of moves is 8), and corrections happen across the moves.

Table \ref{tab:Balance} describes own (panel A) and partner's puzzle behaviors (panel B) and puzzle outcomes (panel C). Panel A shows that there are no gender differences in puzzle-solving ability: both contribution in part 2 and the number of puzzles solved in part 1, the difference between female and male participants are statistically insignificant at 5\% and quantitatively insignificant.\footnote{The number of puzzles solved in part 1 is marginally significant but quantitatively insignificant.}\footnote{I changed the definition of contribution from the one in the pre-analysis plan because there was truncation in the original contribution measure in more than 10\% of the puzzle. Nonetheless, the same results hold when I use original contribution measure; see online Appendix Table \ref{tab:RegH0H1H2SupplOriginalContribution}. Although the original measure is relative to one's pair while the measure I use in this paper is absolute, whether a measure is relative or absolute does not matter because I add individual fixed effects. }\footnote{The correlation coefficient between contribution and number of puzzles solved in part 1 is 0.1059 and the p-value is below 0.001 (with standard errors clustered at individual level).} This is consistent with \textcite{isaksson_it_2018}, who also finds no gender difference in contribution or number of puzzles solved alone using the same puzzle, suggesting that any gender difference I would find is unlikely to come from their ability difference. Panel A also shows that there are no gender differences in propensity to correct partners, suggesting any gender differences I would find are not coming from either gender corrects more than the other gender.

\begin{figure}[!htb]
\begin{center}
\caption{Distribution of contribution}\label{fig:Performance}
\includegraphics[width=\columnwidth]{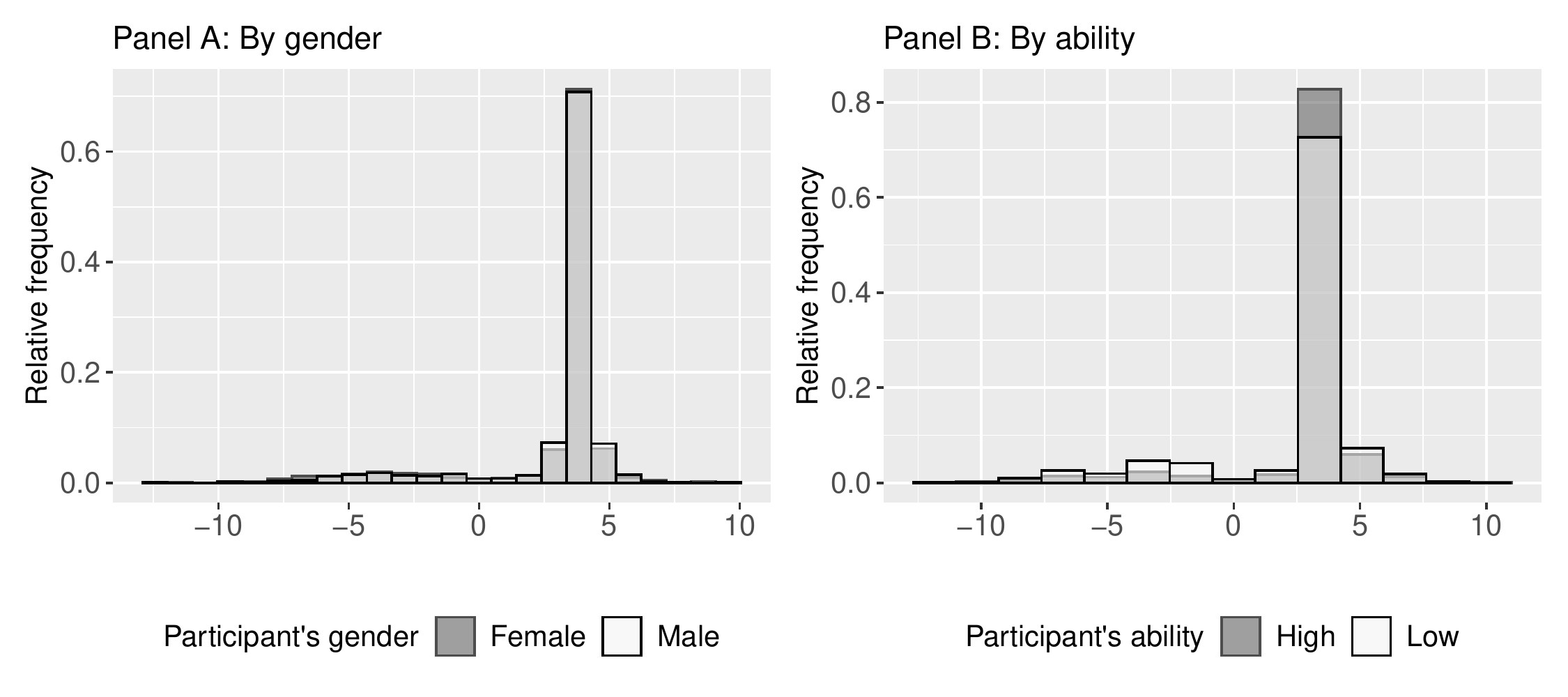}%
\end{center}
\vspace{-12pt}
\begin{tablenotes}[flushleft]
\item \footnotesize \textit{Notes:} This figure shows the distribution of individual contribution by gender (panel A) and ability (panel B) and shows that most participants contributed to the same degree. Panel A further shows no gender difference in contribution, and panel B further shows that high-ability higher fraction of people contributes to the puzzles to the same degree. Contribution is defined as one's net good moves in a given puzzle (the number of good moves minus the number of bad moves).
\end{tablenotes}
\end{figure}

Panel A of Figure \ref{fig:Performance} presents the distribution of contribution by participants gender to further elaborate panel A of Table \ref{tab:Balance} that women and men are equally good at puzzle solving: in about 70\% of the puzzles, participants' contribution is 4 (total good moves minus total bad moves), and women's and men's distributions almost overlap.

Panel B shows that puzzle-solving ability as well as propensity to make corrections (both of a mistake and of a right move) of partners paired with female and male participants is the same, suggesting random pairing was successful and that any gender differences I would find are not coming from partners of either gender correct more often. Participants are corrected by their partner in 15-16\% of the total puzzles, of which 12-13\% are good corrections, and 5-6\% are bad corrections.\footnote{The percentage of good corrections and bad corrections do not sum up to the percentage of any correction means there are puzzles where both good and bad corrections occurred. The results are robust to exclusion of these overlapping puzzles, as shown in online Appendix Figure \ref{fig:RegH0H1H2SupplRobustPlot}.}

Panel C shows that participants state they want to collaborate with the partner 71-72\% of the time. Participants spend on average 43-44 seconds for each puzzle (the maximum time a pair can spend is 120 seconds), and take 11 moves. 85-86\% of the puzzles are solved and participants and the partner correct each other's move consecutively in 4\% of the puzzles.\footnote{Indeed, in puzzles where consecutive correction happens, probability of selecting a paired participant as collaborator drops from 78.0\% to 26.8\%.} There is no gender difference in any of these outcomes, suggesting any gender differences cannot be attributed to the imbalance in these outcomes.\footnote{Note that time spent to solve a puzzle is endogenous to correction and not a good control. For example. if one corrects a mistake, then it takes fewer time to solve the puzzle. If one corrects a right move, on the other hand, then it takes more time to solve the puzzle.}

\section{Theoretical framework}\label{sec:Theory}
I present a simple theoretical framework to provide a benchmark for rational agent's behaviors.

I consider a participant i who maximizes their expected utility by selecting their collaborator j from a set of i's potential collaborators $J\equiv\{1,2,3,4,5,6,7\}$. I assume i can partially observe j's move quality.

i's utility depends on their payoff and emotion. The utility is increasing in the payoff, and the payoff is increasing in i's belief about j's ability. Thus, if i would select with whom to play in part 3, i would face the following problem:
\begin{equation}\label{eq:Theory}
\max_{j\in J} E_{\mu_j}[u_i(\underbrace{\pi(\mu_j(\tilde{a}_j,c_j,f_j))}_{\text{i's payoff}},\underbrace{\kappa_i(c_j,f_j)}_{\text{i's emotion}})|\theta_i,\omega_i],\;\;\; \partial u_i / \partial \pi > 0, \;\partial \pi / \partial \mu_j > 0
\end{equation}
where each term is defined as follows:
\begin{itemize}
\item $\mu_j$: i's belief about j's ability
\item $\tilde{a}_j$: j's ability perceived by i
\item $c_j$: j's good correction (=1 if j corrected i, =0 if j did not correct i)
\item $f_j$: j's gender (=1 if female, =0 if male)
\item $\theta_i$: i's belief about their ability relative to other participants (>0 if high, =0 if same, <0 if low)
\item $\omega_i$: j's belief about women's ability relative to men (>0 if high, =0 if same, <0 if low)
\end{itemize}
I assume:
\begin{itemize}
\item $\mu_j$ is increasing in j's ability perceived by i: $\partial \mu_j / \partial \tilde{a}_j > 0$
\item i's utility is decreasing in their emotion: $\partial u_i / \partial \kappa_i < 0$
\item emotion is irrelevant if i is fully rational: $u_i(\pi,\kappa_i)\propto u_i(\pi)$
\end{itemize}

If i can fully observe j's move quality and i is fully rational, then j's correction, $c_j$, and gender, $f_j$, do not convey any information about j's ability and is irrelevant for i's decision making. However, since i can only partially observe j's move quality, j's correction and gender convey information about j's ability even if i is fully rational.\footnote{I nonparametrically control for j's gender, but I also examine the effect of interaction term between j's correction and j's gender.}

\subsection{When i is fully rational}
First, \textit{keeping j's ability perceived by i fixed}, the information j's correction conveys depends on $\theta_i$. If i believes they are good at the puzzle, they would consider a correction as a signal of low ability because i believes their move is correct. On the other hand, if i believes their ability is low, then they would consider a correction as a signal of high ability. If i believes their ability is the same as j's, then a correction would not convey any information. However, since i can partially observe j's move quality, i consider a good correction as less negative/more positive signal than a bad corrections regardless of $\theta_i$. Thus,
\begin{itemize}
\item $\partial \mu_j / \partial c_j|_{\text{$c_j$ is a bad correction}} < \partial \mu_j / \partial c_j|_{\text{$c_j$ is a good correction}} \;\forall \theta_i$.
\end{itemize}

Similarly, if i believes women is better at the puzzle, they would consider a correction from a woman as a signal of high ability relative to men's correction. On the other hand, if i believes women is worse at the puzzle, then they would consider a correction from a woman as a signal of low ability relative to men's correction. If i believes women and men are equally good at the puzzle, then a correction from a woman or man is irrelevant. Thus,
\begin{itemize}
\item $\partial^2 \mu_j / \partial c_j \partial f_j > 0 \;\forall \theta_i$ if $\omega_i>0$,
\item $\partial^2 \mu_j / \partial c_j \partial f_j > 0 \;\forall \theta_i$ if $\omega_i=0$, and
\item $\partial^2 \mu_j / \partial c_j \partial f_j < 0 \;\forall \theta_i$ if $\omega_i<0$.
\end{itemize}

\subsection{When i is not fully rational}
When i is not fully rational, i's emotion, $\kappa_i$, enters in their maximization problem. Specifically, I assume that j's correction induces i's negative feeling towards j. Also, I assume corrections by women induce i's stronger negative feeling towards j. Thus,
\begin{itemize}
\item $\partial \kappa_i / \partial c_j < 0$ and
\item $\partial^2 \kappa_i / \partial c_j\partial f_j < 0$.
\end{itemize}

\section{Response to corrections}\label{sec:ResponseCorrect}
In this section, I document evidence that people -- both women and men -- understand the notion of good and bad moves. However, they are less willing to work with a person who corrected their move after controlling for that person's contribution to the puzzle, even if that person makes good corrections.

\subsection{Response to corrections: Estimating equation}
I estimate the following model with OLS.
\begin{equation}\label{eq:H0Eq}
\begin{split}
Select_{ij} = \beta_1 CorrectedGood_{ij} + \beta_2 CorrectedBad_{ij} + \beta_3 Female_j + \delta Contribution_j + \mu_i + \epsilon_{ij}
\end{split}
\end{equation}
where each variable is defined as follows:
\begin{itemize}
\item $Select_{ij}\in\{0,1\}$: an indicator variable equals 1 if i selects j as their collaborator, 0 otherwise.
\item $CorrectedGood_{ij}\in\{0,1\}$: an indicator variable equals 1 if j corrected i and moved the puzzle closer to the solution, 0 otherwise.
\item $CorrectedBad_{ij}\in\{0,1\}$: an indicator variable equals 1 if j corrected i and moved the puzzle far away from the solution, 0 otherwise.
\item $Female_j\in\{0,1\}$: an indicator variable equals 1 if j is female, 0 otherwise.
\item $Contribution_j\in \mathbbm{Z}$: j’s contribution to a puzzle played with i.
\item $\epsilon_{ij}$: omitted factors that affect i's likelihood to select j as their collaborator. 
\end{itemize}
and $\mu_i \equiv \sum_{k=1}^N\mu^k \mathbbm{1}[i=k]$ is individual fixed effects, where $N$ is the total number of participants in the sample and $\mathbbm{1}$ is the indicator variable. Standard errors are clustered at the individual level.\footnote{This is because the treatment unit is i. Although the same participant appears twice (once as i and once as j), j is passive in collaborator selection.}

The key identification assumption is that $Contribution_j$ fully captures j's ability \textit{perceived} by i (not true ability).\footnote{By random pairing of participants, the paired participant's gender is exogenous to participant's unobservables. However, correction is not exogenous for two reasons: (i) correction can be correlated with the paired participant's ability and paired participant's ability can affect participant's collaborator selection; (ii) There is an effect similar to the reflection effect: participant's puzzle behavior affects the paired participant's behavior and vice versa; for example, a participant's meanness can increase the paired participant's correction and can also affect their collaborator selection. The identification assumption concerns the former point. To address the latter point, I add individual fixed effects.} This assumption is reasonable if we think participants' willingness to collaborate is increasing in the partner's contribution to the puzzle, which is consistent with that participants can partially observe their partners' ability and their expected utility is increasing in their payoff.

\subsection{Response to corrections: Results}
Columns 1-4 of Table \ref{tab:RegH0H1H2Suppl} present the regression results of equation \ref{eq:H0Eq}. Columns 1 and 3 show that when we do not control for partner's contribution, the coefficient estimate on bad correction is negative and very large: the point estimate is -0.550 (p-value < 0.01) for women and -0.457 (p-value < 0.01) for men. That is, participants are 45.7-55.0\% less willing to collaborate with partners who made a bad correction, a correction that moved the puzzle far away from the solution. Indeed, these coefficient estimates are more negative than the coefficient estimates on good corrections: 0.281 more negative for women (p-value < 0.01) and 0.259 more negative for men (p-value < 0.01). This is evidence that my experimental design is valid: participants correctly understand the notion of good and bad moves and that participants are more willing to collaborate with partners who contributed more.

Looking at columns 2 and 4, the coefficient estimate on the partner's contribution is positive and quantitatively and statistically highly significant: the point estimate is 0.089 (p-value < 0.01) for women and 0.080 (p-value < 0.01) for men. This suggests that participants are 8.0-8.9\% more willing to collaborate with partners who make one more good move.

The coefficient estimates on good correction in columns 2 and 4 are negative and quantitatively and statistically highly significant with the point estimate -0.229 (p-value < 0.01) and -0.168 (p-value < 0.01). This suggests that people are 16.8-22.9\% less willing to collaborate with those who made a good correction(s), which corresponds to an increase in the contribution by 0.72-0.93 standard deviation.\footnote{The standard deviation is taken from panel B of Table \ref{tab:Balance}: 2.73 for partners faced by women 2.87 for and partners faced by men.}

The coefficient estimate on bad correction in columns 2, which shows women's response, is negative and quantitatively and statistically highly significant with the point estimate of -0.172 (p-value < 0.01). However, the corresponding coefficient estimate for men in column 4 is -0.011 and statistically insignificant even at 10\%. This may indicate that there are no emotional factors associated with bad corrections for men, but the number of good and bad corrections are not large enough, and I may be picking up some irregularities in the data.

\section{Do women's corrections receive stronger negative reactions?}\label{sec:ResponseByCorrectorGender}
In this section, I document that neither men nor women underestimate women's contribution and that women's corrections do not receive stronger negative reactions by either women or men.

\subsection{Do women's corrections receive stronger negative reactions? Estimating equation}
I estimate the following model with OLS.
\begin{equation}\label{eq:H1H2Eq}
\begin{split}
Select_{ij} = &\beta_1 CorrectedGood_{ij} + \beta_2 CorrectedBad_{ij} + \beta_3 Female_j \\&+ \beta_4 CorrectedGood_{ij} \times Female_j + \beta_5 CorrectedBad_{ij} \times Female_j \\&+ \delta_1 Contribution_j + \delta_2 Contribution_j\times Female_j + \mu_i + \epsilon_{ij}
\end{split}
\end{equation}
Where each variable is defined as in equation \ref{eq:H0Eq}.

\subsection{Do women's corrections receive stronger negative reactions? Results}
Columns 5-6 of Table \ref{tab:RegH0H1H2Suppl} present the regression results of equation \ref{eq:H1H2Eq}. First, in both columns, the coefficient estimate on the interaction between partner's contribution and female partner is almost 0 and statistically insignificant even at 10\%. This suggests that neither women nor men underestimate women's contribution when selecting a collaborator.

In column 5, the coefficient estimates on the interaction between good correction and female partner and bad correction and female partner are all positive although statistically insignificant. This suggests that, if anything, women respond slightly less negatively to women's correction. In column 6, however, the coefficient estimate on the interaction between good correction and female partner is negative and statistically marginally significant at 10\%. Yet, it is not statistically significant at 5\%, and that the coefficient estimate on the interaction between bad correction and female partner is positive on the contrary, although statistically insignificant. Thus, women's corrections do not receive stronger negative reactions from either women or men.

\section{Who respond negatively to corrections?}\label{sec:ResponseByAbility}
If only low-ability people respond negatively to corrections, then correcting colleagues may not be very costly. However, in this section, I document that even high-ability people respond negatively to corrections.

\subsection{Who respond negatively to corrections? Estimating equation}
I estimate the following model with OLS.
\begin{equation}\label{eq:SupplEq}
\begin{split}
Select_{ij} = &\beta_1 CorrectedGood_{ij} + \beta_2 CorrectedBad_{ij} + \beta_3 Female_j \\&+ \beta_4 CorrectedGood_{ij} \times HighAbility_i + \beta_5 CorrectedBad_{ij} \times HighAbility_i \\&+ \delta_1 Contribution_j + \delta_2 Contribution_j\times HighAbility_i + \mu_i + \epsilon_{ij}
\end{split}
\end{equation}
where each variable is defined as follows:
\begin{itemize}
\item $HighAbility_i\in\{0,1\}$: an indicator variable equals 1 if i solved the above-median number of puzzles in part 1 in a session they have participated, 0 otherwise.
\end{itemize}
Other variables are as defined in equations \ref{eq:H0Eq}.

\subsection{Who respond negatively to corrections? Results}
Columns 7-8 of Table \ref{tab:RegH0H1H2Suppl} present the regression results of equation \ref{eq:SupplEq}. First, in both columns, the coefficient estimate on the interaction between partner's contribution and female partner is almost 0 and statistically insignificant even at 10\%. This may seem counter-intuitive because high-ability people should be better able to observe move quality. One explanation is that because high-ability people are paired with high-ability people, so the distribution of contribution is less dispersed and contribution plays weaker role in willingness to collaborate. Panel B of Figure \ref{fig:Performance} shows distribution of contribution of high-ability people is indeed less dispersed.

In column 7, the coefficient estimate on the interaction between good correction and high-ability is negative and the coefficient estimate on the interaction between bad correction and high-ability is positive, but they are both statistically insignificant. This suggests that high-ability women respond as negatively to corrections as low-ability women.

In column 8, the coefficient estimates on the interaction between good correction and high-ability and bad correction and high-ability are both negative and the former is statistically significant at 5\%. This is hard to interpret -- it may be capturing some irregularity in the data -- but the bottom line is that high-ability men also respond as negatively to corrections as low-ability men.

\section{Discussion}\label{sec:ExtValid}
This paper demonstrates that people, including those with high productivity, are less willing to collaborate with a person who has corrected them even if the correction improves group performance. However, I do not find evidence that people respond more negatively to corrections by women. Thus, although women do not face a higher hurdle in their career, correcting colleagues is costly and reduces group efficiency.

While the laboratory setting is different from the real-world, my findings are likely to be a lower bound because of the following three reasons. First, being corrected is not observed by others in my experiment: those who have been corrected do not face any reputation cost, unlike in the real-world. Second, the emotional stake is much smaller: the puzzle ability is not informative of the ability relevant for their work or study; it is not something people have been devoting much of their time to, such as university exams, academic research, and corporate investment projects. Third, participants are equal in my experiment; in the real-world, on the other hand, there are sometimes senior-junior relationships, and corrections by junior people may induce stronger negative reactions.

But there are two caveats. The first is that participants are strangers to each other in my experiments while people know each other in the real-world. Thus, it is possible that repeated interactions would mitigate people's negative response to corrections (but they may also magnify the negative response due to rivalry, failure to build a good rapport, etc.). The second is that most participants are bachelor or master students who are supposed to have a weaker gender bias. Women's corrections may receive stronger negative reactions if participants are older.

\newpage
\clearpage
\begin{singlespace}
\printbibliography
\end{singlespace}

\newpage
\clearpage
\section*{Figures}

\newpage
\clearpage
\section*{Tables}

\begin{table}[!htb]
\begin{center}
\caption{Own and partners' puzzle behaviors and puzzle outcomes}\label{tab:Balance}
\begin{adjustbox}{max width=\textwidth}
\input{Balance.tex}

\end{adjustbox}
\end{center}
\vspace{-12pt}
\begin{tablenotes}[flushleft]
\item \footnotesize \textit{Notes:} This table describes own (panel A) and partner's puzzle behaviors (panel B) and puzzle outcomes (panel C). P-values of the difference between female and male participants are calculated with standard errors clustered at the individual level. Contribution is defined as one's net good moves in a given puzzle (the number of good moves minus the number of bad moves).
\end{tablenotes}
\end{table}%

\begin{table}[!htb]
\begin{center}
\caption{Response to corrections}\label{tab:RegH0H1H2Suppl}
\begin{adjustbox}{max width=\textwidth}
\input{RegH0H1H2Suppl.tex}
\end{adjustbox}
\end{center}
\vspace{-12pt}
\begin{tablenotes}[flushleft]
\item \footnotesize \textit{Notes:} Columns 1 and 3 presents the regression results of equation \ref{eq:H0Eq} without partners' contribution and shows that people -- both women and men -- understand the notion of good and bad moves. Columns 2 and 4 show that they are less willing to work with a person who corrected their move after controlling for that person's contribution to the puzzle, even if that person makes good corrections. Columns 5 and 6 show neither men nor women underestimate women's contribution and that women's corrections do not receive stronger negative reactions by either women or men. Columns 7 and 8 show even high-ability people respond negatively to corrections. \sdnote. \starnote.
\end{tablenotes}
\end{table}%

\end{document}

%% file: Balance.tex
\begin{tabular}{lccccccc}
\toprule
 & \multicolumn{2}{c}{\thead{Female \\(N=1708)}} & \multicolumn{2}{c}{\thead{Male \\(N=1540)}} & \multicolumn{3}{c}{\thead{Difference \\ (Female -- Male)}}  \\
 & Mean & SD & Mean & SD & Mean & SE & P-value\\
\midrule
\multicolumn{8}{l}{\underline{Panel A: Own behaviors}} \\
Contribution & 2.98 & 2.93 & 3.14 & 2.64 & -0.16 & 0.10 & 0.11\\
\# puzzles solved in pt. 1 & 8.36 & 2.41 & 8.80 & 2.34 & -0.44 & 0.22 & 0.05\\
Correction & 0.15 & 0.36 & 0.16 & 0.36 & 0.00 & 0.01 & 0.85\\
Good correction & 0.12 & 0.33 & 0.12 & 0.33 & 0.00 & 0.01 & 0.90\\
Bad correction & 0.06 & 0.23 & 0.05 & 0.22 & 0.00 & 0.01 & 0.70\\
\midrule
\multicolumn{8}{l}{\underline{Panel B: Partner's behaviors}} \\
Contribution & 3.04 & 2.73 & 3.07 & 2.87 & -0.03 & 0.10 & 0.77\\
\# puzzles solved in pt. 1 & 8.58 & 2.35 & 8.57 & 2.43 & 0.01 & 0.16 & 0.93\\
Correction & 0.16 & 0.37 & 0.15 & 0.36 & 0.01 & 0.01 & 0.51\\
Good correction & 0.13 & 0.33 & 0.12 & 0.32 & 0.01 & 0.01 & 0.44\\
Bad correction & 0.06 & 0.23 & 0.05 & 0.22 & 0.01 & 0.01 & 0.44\\
\midrule
\multicolumn{8}{l}{\underline{Panel C: Puzzle outcomes}} \\
Willing to collaborate (yes=1, no=0) & 0.72 & 0.45 & 0.71 & 0.45 & 0.01 & 0.02 & 0.49\\
Time spent (sec.) & 43.74 & 36.15 & 42.99 & 35.76 & 0.74 & 1.28 & 0.56\\
Total moves & 11.18 & 7.46 & 11.21 & 7.70 & -0.03 & 0.28 & 0.92\\
Puzzle solved & 0.85 & 0.36 & 0.86 & 0.35 & -0.01 & 0.01 & 0.43\\
Consecutive correction & 0.04 & 0.20 & 0.04 & 0.21 & 0.00 & 0.01 & 0.81\\
\bottomrule
\end{tabular}%

%% file: RegH0H1H2Suppl.tex
\begin{tabular}{lcccccccc}
\toprule
Outcome: & \multicolumn{8}{c}{Willing to collaborate (yes=1, no=0)} \\
\midrule
Sample: & \multicolumn{2}{c}{Female} & \multicolumn{2}{c}{Male} & Female & Male & Female & Male \\
  & (1) & (2) & (3) & (4) & (5) & (6) & (7) & (8) \\
\midrule
Good correction & -0.269*** & -0.229*** & -0.197*** & -0.168*** & -0.248*** & -0.104* & -0.208*** & -0.107***\\
 & (0.043) & (0.033) & (0.040) & (0.036) & (0.045) & (0.053) & (0.042) & (0.041)\\
Bad correction & -0.550*** & -0.172*** & -0.457*** & -0.011 & -0.218*** & -0.104 & -0.201*** & 0.005\\
 & (0.044) & (0.047) & (0.050) & (0.052) & (0.064) & (0.076) & (0.064) & (0.063)\\
Female partner & -0.009 & 0.004 & 0.007 & 0.016 & -0.002 & 0.003 & 0.002 & 0.014\\
 & (0.021) & (0.018) & (0.026) & (0.021) & (0.032) & (0.030) & (0.018) & (0.021)\\
Partner's contribution &  & 0.089*** &  & 0.080*** & 0.089*** & 0.077*** & 0.089*** & 0.082***\\
 &  & (0.004) &  & (0.004) & (0.006) & (0.006) & (0.005) & (0.004)\\
Good correction x Female partner &  &  &  &  & 0.035 & -0.119* &  & \\
 &  &  &  &  & (0.057) & (0.067) &  & \\
Bad correction x Female partner &  &  &  &  & 0.090 & 0.168 &  & \\
 &  &  &  &  & (0.093) & (0.102) &  & \\
Partner's contribution x Female partner &  &  &  &  & -0.001 & 0.006 &  & \\
 &  &  &  &  & (0.008) & (0.007) &  & \\
Good correction x High ability &  &  &  &  &  &  & -0.048 & -0.180**\\
 &  &  &  &  &  &  & (0.066) & (0.075)\\
Bad correction x High ability &  &  &  &  &  &  & 0.074 & -0.061\\
 &  &  &  &  &  &  & (0.095) & (0.109)\\
Partner's contribution x High ability &  &  &  &  &  &  & -0.001 & -0.003\\
 &  &  &  &  &  &  & (0.007) & (0.008)\\
\midrule
Individual FE & \cmark & \cmark & \cmark & \cmark & \cmark & \cmark & \cmark & \cmark \\
\midrule
\multirow{2}{*}{\makecell[l]{Good correction\\$-$Bad correction}} & 0.281*** & -0.057 & 0.259*** & -0.157** &  &  &  & \\
 & (0.075) & (0.061) & (0.071) & (0.065) &  &  &  & \\
Baseline mean & 0.780 & 0.780 & 0.778 & 0.778 & 0.780 & 0.778 & 0.780 & 0.778\\
Baseline SD & 0.414 & 0.414 & 0.416 & 0.416 & 0.414 & 0.416 & 0.414 & 0.416\\
Adj. R-squared & 0.111 & 0.369 & 0.090 & 0.306 & 0.369 & 0.307 & 0.368 & 0.308\\
Observations & 1670 & 1670 & 1510 & 1510 & 1670 & 1510 & 1670 & 1510\\
Clusters & 244 & 244 & 220 & 220 & 244 & 220 & 244 & 220\\
\bottomrule
\end{tabular}